\documentstyle[12pt]{article}
\newcommand{\be}{\begin{equation}}
\newcommand{\ee}{\end{equation}}
\newcommand{\bea}{\begin{eqnarray}}
\newcommand{\eea}{\end{eqnarray}}
\newcommand{\vte}{{\mbox{\boldmath$T$}}}
\newcommand{\vs}{{\mbox{\boldmath$S$}}}
\newcommand{\vt}{{\mbox{\boldmath$\tau$}}}
\newcommand{\vvs}{{\mbox{\boldmath$\sigma$}}}
\newcommand{\veta}{{\mbox{\boldmath$\eta$}}}
\def\lr#1{\langle#1\rangle}

\advance\hoffset -5mm
\begin{document}
\parindent 20pt
\baselineskip 7mm
\pagestyle{plain}
\pagenumbering{arabic}
\setcounter{page}{1}
\date{}
\title{Multi--Channel Kondo Necklace}
\author{P. Fazekas\thanks{Permanent address: Research Institute for Solid
State Physics, P.O.B. 49, Budapest 114, H--1525 Hungary}\ \ and Hae--Young
Kee\thanks{Permanent address:
Department of Physics Education, Seoul National Unversity, Seoul, 151--742
Korea}\\
International Centre for Theoretical Physics,\\
P.O. Box 586, I--34100 Trieste, Italy}

\maketitle

\vskip 1 cm

\begin{abstract}
A multi--channel generalization of Doniach's Kondo necklace model is
formulated, and its phase diagram studied in the mean--field approximation.
Our intention is to introduce the possible simplest model which displays some
of the features expected from the overscreened Kondo lattice.
The $N$ conduction electron channels are represented by $N$ sets of
pseudospins $\vt_{j}$, $j=1, ... , N$, which are all antiferromagnetically
coupled to a periodic array of $|\vs|=1/2$ spins. Exploiting permutation
symmetry in the channel index $j$ allows us to write down the
self--consistency equation for general $N$. For $N>2$, we find that the
critical temperature is rising with increasing Kondo interaction; we interpret
this effect by pointing out that the Kondo coupling creates the composite
pseudospin objects which undergo an ordering transition. The relevance of our
findings to the underlying fermionic multi--channel problem is discussed.
\end{abstract}

\newpage

\section{\normalsize INTRODUCTION}

  The essential physics of mixed valent and heavy fermionic systems
\cite{rev} is habitually described by some suitable version of the periodic
Anderson model \cite{lrs}. In this framework, heavy fermionic systems
correspond to the limiting case of nearly integral valence, which permits us
to use a Kondo lattice model for the approximate handling of the
low--energy, low--temperature behaviour of the Anderson lattice \cite{klm}.
Though already several
steps removed from physical reality, it is widely accepted that the Kondo
lattice model incorporates the basic ingredients necessary to understand the
competition between two opposing tendencies: the formation and eventual
ordering of magnetic moments, and the formation of a heavy Fermi sea.
Research has for a long time
been  focussed on the case where localized spins $S=1/2$ are compensated by
electrons moving in a non--degenerate band
\\
\be
H_{KL} = \sum_{{\bf k},\sigma}\epsilon({\bf k})c_{{\bf k}\sigma}^{+}
c_{{\bf k}\sigma} + \frac{J}{2L}\sum_{\bf g}\sum_{{\bf k},{\bf k}^{\prime}}
e^{i({\bf k}-{\bf k}^{\prime}){\bf g}} \sum_{\alpha\beta}
c_{{\bf k}\alpha}^{+}\vvs_{\alpha\beta}c_{{{\bf k}^{\prime}}\beta} \vs
\ee
\\
where the first term describes a non--degenerate band, and the second the
antiferromagnetic ($J>0$) Kondo coupling at each site ${\bf g}$. $L$ is the
number of lattice sites, and $\vvs$ is the vector of Pauli matrices.
$H_{KL}$ is the simplest version of the Kondo lattice hamiltonian; we expect
that it shows the same overall features as the $S=N/2$ models, where
$N$ is the number of conduction electron channels participating in the
Kondo screening process. This special case of exact compensation seems to be
the most favourable for the formation of an overall singlet, i.e., a heavy
Fermi liquid without any residue of magnetic order. Even these perfectly
screened models are, of course, capable of magnetic ordering
\cite{{klm},{lc},{FMh},{DF}}, but the arising of a completely
non--magnetic state can at least be claimed to be a natural option
\cite{ShF}.

In recent years, it has become increasingly clear that considering the
underscreened $S>N/2$, and overscreened $S<N/2$, models brings not just
additional complications but interesting new physics, and it is also
essential
to describe many, if not all, of the actual heavy fermionic materials.

The properties of an isolated underscreened, or overscreened, Kondo impurity
\cite{NoBl} have been studied in great detail, and for a number of physical
quantities, exact results are known \cite{{AD},{WTs}}. Still, understanding
can not yet be considered complete; in particular, we could wish for a
clearer physical picture  of the low--temperature behaviour of the
overscreened model with its intriguing non--integer ground state degeneracy
\cite{{AL91},{GAC93}}.

  Much less is known about magnetic ordering in the corresponding
  {\em periodic models}. Just as in the case  of perfect screening
$S=N/2$, there must be a competition between Kondo screening and intersite
interactions; however, the picture is complicated by the fact that the local
object which tends to arise from the Kondo effect, has a more intricate
internal structure. For the underscreened case, even a very strong
Kondo coupling can not do better than to leave a reduced spin $S-N/2$; thus
one is led to consider the ordering of the residual magnetic moments. The
underscreened Kondo lattice is apparently destined to become magnetic
\cite{CG}, and it was suggested that its study should lead to understanding
heavy fermion magnetism in general \cite{GCA92}.

  The situation is much less clear in the overscreened case. The model which
was first introduced as a matter of academic interest \cite{NoBl}, was
later argued to represent the physical situation for two--level systems
interacting with several conduction electron channels \cite{{Z},{MuGu}}, and
for U--based heavy fermion systems \cite{Cox87}. Subsequent experiments
on structurally disordered metallic nanoconstrictions \cite{RBu}, and on the
Kondo alloy Y$_{1-x}$U$_x$Pd$_3$
\cite{Sea}, gave ample confirmation of the applicability of the overscreened
Kondo model. Another remarkable realization of the multi--channel Kondo
effect is provided by the transport properties of  Pb$_{1-x}$Ge$_x$Te
\cite{KMF}.--- The solution of
the single--impurity problem shows a puzzling mixture of features which one
would naively associate either with the absence, or the completion, of spin
compensation. The ground state possesses a zero point entropy
\cite{{WTs},{AL91},{GAC93}} which, in the limit of an infinite Fermi sea,
does not correspond to a half--integer spin, but reduces to $\ln{(2)}$ if
the system is finite. The finding of a zero point entropy might lead us to
think that the ground state is possessing a residual magnetic moment;
however, calculations show \cite{{AD},{WTs}} that magnetism
is quenched as $T{\to}0$, albeit much more slowly than in the $S=N/2$
case. All in all, it turns out to be fairly difficult to form an intuitive
picture of the multichannel Kondo system whose behaviour is governed by an
intermediate--coupling fixed point. Correspondingly, we have only the
haziest idea of what a periodic array of overscreened Kondo centres is
supposed to be doing.

In either case, when we undertake the study of the magnetic phases of the
periodic models, we have to deal with the ordering of composite spins
created by a local Kondo coupling; and the degree to which Kondo compensation
can progress is itself limited by the intervention of intersite interactions.
Curiously, the effect can go opposite ways in the underscreened, and
overscreened, cases. For the underscreened lattice, the resulting total spin
is necessarily less than $S$, so the ordering temperature is found \cite{FK}
to show a decreasing tendency with increasing $J$. In the overscreened case,
the Kondo effect is actually building up a screening object which can be
ultimately larger than the screened spin $S$. We are going to find this
tendency for sufficiently large ($N\ge 3$) numbers of channels.

\section{\normalsize PSEUDOSPIN MODELS: THE KONDO NECKLACE AND ITS
GENERALIZATIONS}

The fact that the conduction electrons are forming a Fermi sea, with a large
number of arbitrarily low--lying excitations, is essential for the
appearance of a non--analytic energy scale [the Kondo temperature
$T_{K}\propto B\exp{(-1/J\rho(\epsilon_{F}))}$, where $B$ is of the order of
the bandwidth, and $\rho(\epsilon_{F})$ the density of states at the Fermi
level] in the
theory of the single--ion Kondo effect. It is, as yet, an open question
whether there is a non--analytic energy scale associated with the formation
of a non--magnetic ground state of the periodic Kondo model (or, to put it
more pointedly, whether there is a Kondo--effect in the Kondo lattice
\cite{Sch}),
and if yes, whether the ``lattice Kondo temperature'' is different from the
single--ion $T_K$ \cite{{FMh},{ShF},{RU},{Tsu}}. It can be, however, argued
\cite{LiCo} that the competition between the Kondo effect, and
the interactions opposing it, can be successfully mimicked by just
considering the gross effects of different couplings: one can have a crude
version of the phase diagram of the Kondo lattice without the ``true'' Kondo
effect. One can then forget about the low--lying electron--hole excitations
which can lead to infrared divergencies, and consider the conduction
electrons only to the extent that they provide spins which tend to be aligned
antiparallel to the localized spins. One can thus reduce the mixed
spin--fermion problem to a pure spin problem. This was the rationale
behind introducing Doniach's \cite{Do77} necklace model
\\
\be
H_{KN} =  J\sum_{\bf m}\vs_{\bf m}\cdot\vt_{\bf m} +
W \sum_{\lr{\bf mn}} (\tau_{\bf m}^{x}\tau_{\bf n}^{x} +
\tau_{\bf m}^{y}\tau_{\bf n}^{y})
\ee
\\
where the pseudospins $|\vt|=1/2$ represent the spin degrees of freedom of
the conduction electrons, and $W$ is a characteristic amplitude of their
propagation. $H_{KN}$ is the pseudospin version of the $S=1/2$
lattice model given in eqn. (1); clearly, $W$ is an effective parameter which
depends in a complicated manner on the parameters of the underlying electronic
model. The mean field treatment of $H_{KN}$ by Doniach \cite{Do77}
led to a ground state
phase diagram which looks qualitatively the same as expected from a (still
missing) complete treatment of (1), for the case of a half--filled
conduction band: the ground state is antiferromagnetic
for $J/W<1$, and non--magnetic (apparently Kondo--compensated) for $J/W>1$.
We conclude that the study of necklace--type pseudospin models is a useful
prelude to the investigation of the full fermionic Kondo lattice problems. ---
We note that recently, Strong and Millis \cite{Smi} investigated a
generalized form of the $S=1/2$, $N=1$ necklace model with the ultimate
purpose of gaining insight into the behaviour of heavy fermionic systems.

We should emphasize that we use the term ``necklace'', as opposed to
``lattice'',
models to signify that the conduction electron sea has been replaced by a
set of pseudospins, and {\em not} to mean that the model is necessarily
one--dimensional. In (2), the sum over $\lr{\bf mn}$ is over nearest--neighbour
(nn) pairs in any lattice. In fact, usually we will resort to the mean field
approximation (MFA), so lattice dimensionality plays no essential role. The
results obtained in the MFA are thought to be a reasonable approximation for
the three-dimensional models, but they should serve as a rough guide even in
one--dimension \cite{SaSo}, except that whenever long--range order is found,
it should be interpreted as indication of quasi--long--range order.

Considering the rudimentary stage of the understanding of the underscreened,
and quite particularly, the overscreened, periodic Kondo models, we can
appreciate the need for any insight to be gained from the study of related
simpler models. In \cite{FK}, we introduced the underscreened, spin--$S$
necklace models, and discussed their properties. The underscreened necklace
models turn out to be inherently antiferromagnetic, with an easy--plane
anisotropy; the Kondo coupling merely influences the size of the ordered
moment. The relative ease of the solution is thanks to the fact that the
dimensionality of the Hilbert space is merely $\propto S$.

In the present work, we perform a similar investigation of a class of
{\em overscreened necklace models}. We introduce the hamiltonian
\\
\be
H = \frac{J}{N} \sum_{\bf m}\vs_{\bf m}\cdot\sum_{j=1}^{N}\vt_{j,{\bf m}} +
\frac{W}{N} \sum_{\lr{\bf mn}}\sum_{j=1}^{N}
(\tau_{j,{\bf m}}^{x}\tau_{j,{\bf n}}^{x} +
\tau_{j,{\bf m}}^{y}\tau_{j,{\bf n}}^{y})
\ee
\\
where $N$ different kinds of pseudospins $\vt_{j}$ for $j=1,...,N$ are meant
to represent the $N$ screening channels. The localized moments (in truth,
$f$--spins) are taken to be $|\vs|=1/2$. The $x-y$ coupling which should
correspond to the propagating character of the conduction electron spin
degrees of freedom, acts independently in each channel. In order to have a
meaningful large--$N$ limit of the model, one has to keep the individual
couplings of order $1/N$. We are restricting ourselves to the
case when all channels are equivalent; one should be aware, though, that
models with inequivalent channels \cite{Cox1} should be of great physical
interest. The study of some such models is in progress.

A general comment about the necklace replacements of the true Kondo problems
should be made here: in the necklace models, the Kondo effect, and the Kondo
bound state, become very local. This corresponds to the actual physical
situation for very strong Kondo couplings. Thus, the necklace models can be
used in good faith for $S{\ge}k/2$, where the large--$J$ limit is
continuously connected to the regime of small $J$'s. The finding of an
intermediate coupling fixed point
for $S<k/2$ tells us that this is assuredly not the case for an overscreened
Kondo impurity. However, we would still like to argue that the study of the
lattice model (3) should have relevance for the overscreened
periodic Kondo model. It can be generally argued  that in Kondo
lattice models, especially in the presence of ordering, intersite effects
may prevent us from reaching the fixed point
of the corresponding single--site models, and so the subtleties of the
impurity solution may become irrelevant for the lattice. This argument
should apply quite particularly to the overscreened case where the behaviour
is known to be extremely sensitive to the presence of external fields
\cite{{WTs},{SaS91}}. It has been pointed out by Andraka and Tsvelik
\cite{ATs} that intersite interactions, acting like fields, can be expected
to turn the system towards a strong--coupling fixed point. An important
piece of corroborative evidence is that for the two--impurity problem,
destabilization
of the ``marginal Fermi liquid'' single--impurity behaviour was found by
Ingersent, Jones and Wilkins \cite{IJW}. Thus we think it
not implausible that in periodic models, ordering of tightly bound objects
created by the multichannel Kondo coupling can be taking place, and then our
model hamiltonian (3) acquires relevance.
\\

\section{\normalsize MEAN--FIELD THEORY OF THE OVERSCREENED NECKLACE MODEL}

The overscreened model presents much larger difficulties than the
underscreened one \cite{FK} because the dimensionality of the Hilbert space
is now exponentially large in $N$. We are, however, aided by the huge
symmetry of $H$: neglecting lattice translations, the symmetry
group is $G=SO(3){\bigotimes}S_{N}$. $H$ has full spin--rotational symmetry,
as well as invariance under arbitrary permutations of the channel indices $j$.
We will find that symmetry simplifies the appearance of the spectrum to such
an extent that at least the single--site mean field problem becomes solvable
for general $N$, without any further approximations.

Assuming a simple magnetic order in which all neighbours of a given site have
their $\tau$-spins polarized in the same direction, we can do the familiar
MF decoupling to arrive at the single--site hamiltonian
\\
\be
H_{MF} = \frac{J}{N} \vs\cdot\sum_{j=1}^{N}\vt_{j} -
\frac{\omega}{N}\sum_{j=1}^{N}\tau_{j}^{z}
\ee
\\
where $\omega=Wz\lr{\tau}$ is the mean field strength, $z$ being the
coordination number. For the sake of solving
just the MF problem, we rotated the quantization axis to align up with the
mean field; in the sense of the full hamiltonian $H$, this must be an arbitrary
direction in the $x-y$ plane. For bipartite lattices, there is no formal
difference between ferromagnetic $W<0$, and antiferromagnetic $W>0$, intersite
couplings; to have a closer correspondence with physical reality, we will have
to take $W>0$.

The Kondo term in $H_{MF}$ has the single--site version of the high symmetry
described before:
it is invariant under arbitrary rotations of the total spin $\veta=\vs+
\sum_{j=1}^{N}\vt_{j}$, as well under permutations of the $N$ channels.
Switching on the effective field destroys spin--rotational invariance, but we
will still find it useful to think of the states as derived from splitting
the highly degenerate levels of the underlying $SO(3)$--invariant problem.
In any case, we have still got the permutational symmetry. Actually, this
rests on having chosen a $j$--independent effective field:
\\
\be
\lr{\tau_{1}^{z}}= ... =\lr{\tau_{N}^{z}}=\lr{\tau}
\ee
\\
In principle, MF solutions breaking the $S_{N}$ symmetry are imaginable.
For small values of $N$, we were exploring this possibility, and found that
the symmetrical solution is more advantageous, so we feel confident in using
the form of $H_{MF}$ specified in (4).

In what follows, we set out to classify the eigenstates according to the
irreducible representations of $S_N$, by constructing their Young--tableaux
\cite{group}.
As it turns out, this classification is intimately related to the
classification according to the total spin $\eta$, and its $z$--component
$\eta^z$, so it is useful to start with the high--symmetry problem $\omega=0$.

Let us start with the unique $\eta^{z}=(N+1)/2$ state
\\
\setlength{\unitlength}{1cm}
\begin{equation}
\begin{picture}(5,2)(0,0)
\thicklines
\put(1.0,0.0){\line(1,0){4.0}}
\put(1.0,1.0){\line(1,0){4.0}}
\put(1.0,1.0){\line(0,-1){1.0}}
\put(2.0,1.0){\line(0,-1){1.0}}
\put(4.0,1.0){\line(0,-1){1.0}}
\put(5.0,1.0){\line(0,-1){1.0}}
\put(1.5,0.5){\makebox(0,0)[c]{$\uparrow$}}
\put(2.0,0.5){\makebox(2,0)[c]{...}}
\put(4.5,0.5){\makebox(0,0)[c]{$\uparrow$}}
\put(0.5,0.5){\makebox(0,0)[c]{\boldmath {$\Uparrow$}}}
\put(3.0,1.3){\oval(3.7,.3)[t]}
\put(3.0,1.8){\makebox(0,0)[c]{\it N}}
\end{picture}
\end{equation}
\\
where {\boldmath {$\Uparrow$}} denotes the $S$--spin, and the $\tau$--spins
enter the Young tableau (YT) for the identity representation of $S_N$.
Obviously, the $S$--spin cannot be included in the YT because there is no
symmetry with respect to interchanging $\vs$ with one of the $\vt$s.
Following custom, we will refer to the $\tau$--part of such a graphical
representation as a YT, while the whole picture will be called a ``diagram'';
since with the inclusion of the $S$--spin, it is not a YT in the
conventional sense.

(6) is the $\eta^{z}=(N+1)/2$ component of the maximum spin
$\eta=(N+1)/2$ multiplet. Its energy is
\\
\be
\epsilon_{0}^{+} = \frac{J}{4}
\ee
\\

In the subspace $\eta^{z}=(N-1)/2$ we are going to meet several kinds of
states. The permutationally symmetrical ones are
\\
\setlength{\unitlength}{1cm}
\begin{equation}
\begin{picture}(5,2)(0,0)
\thicklines
\put(1.0,0.0){\line(1,0){4.0}}
\put(1.0,1.0){\line(1,0){4.0}}
\put(1.0,1.0){\line(0,-1){1.0}}
\put(2.0,1.0){\line(0,-1){1.0}}
\put(4.0,1.0){\line(0,-1){1.0}}
\put(5.0,1.0){\line(0,-1){1.0}}
\put(1.5,0.5){\makebox(0,0)[c]{$\uparrow$}}
\put(2.0,0.5){\makebox(2,0)[c]{...}}
\put(4.5,0.5){\makebox(0,0)[c]{$\uparrow$}}
\put(0.5,0.5){\makebox(0,0)[c]{\boldmath {$\Downarrow$}}}
\put(3.0,1.3){\oval(3.7,.3)[t]}
\put(3.0,1.8){\makebox(0,0)[c]{\it N}}
\end{picture}
\end{equation}
\\
and
\\
\begin{equation}
\begin{picture}(6,2)(0,0)
\thicklines
\put(1.0,0.0){\line(1,0){5.0}}
\put(1.0,1.0){\line(1,0){5.0}}
\put(1.0,1.0){\line(0,-1){1.0}}
\put(2.0,1.0){\line(0,-1){1.0}}
\put(4.0,1.0){\line(0,-1){1.0}}
\put(5.0,1.0){\line(0,-1){1.0}}
\put(6.0,1.0){\line(0,-1){1.0}}
\put(1.5,0.5){\makebox(0,0)[c]{$\uparrow$}}
\put(2.0,0.5){\makebox(2,0)[c]{...}}
\put(4.5,0.5){\makebox(0,0)[c]{$\uparrow$}}
\put(5.5,0.5){\makebox(0,0)[c]{$\downarrow$}}
\put(0.5,0.5){\makebox(0,0)[c]{\boldmath {$\Uparrow$}}}
\put(3.0,1.3){\oval(3.7,.3)[t]}
\put(3.0,1.8){\makebox(0,0)[c]{\it N-1}}
\end{picture}
\end{equation}
\\
As usual, YT represent states fully symmetrized along a row, so it is
unambiguous what (8), and (9), stand for. These are not eigenstates of the
total Kondo coupling; the actual eigenstates are their properly chosen
linear combinations. One of these must be the $\eta^{z}=(N-1)/2$ component
of the $\eta=(N+1)/2$ multiplet. The other belongs to a symmetrical
$\eta^{z}=(N-1)/2$ multiplet, with the energy
\\
\be
\epsilon_{0}^{-} = -\frac{N+2}{4N}J
\ee
\\
This is actually the ground state energy of the local Kondo coupling.
The corresponding eigenstate can be roughly described by saying that the
$\vs$--spin is pointing antiparallel to an assembly of $\vt$--spins.
Actually, it is not quite that: it contains an admixture of states (9) in
which a $\vt$--spin is antiparallel to all other spins.

The ground state of the Kondo--term is $N$--fold degenerate. We can interpret
this as a result of an effective ferromagnetic coupling which the common
Kondo--coupling to the $\vs$--spin induces between the $\vt$--spins.
This feature is not unlike to that found in rigorous treatments of the
overscreened Kondo impurity \cite{{WTs},{SaS91}}. --- Note, however, that
here it leads to ascribing the zero--point entropy $k_{B}\ln{N}$ to a Kondo
site and thus , as discussed in the Introduction, we fail to recover the
subtle features of the single--ion solution of the original overscreened
Kondo problem. However, we have the correct features for a strong--coupling
solution which may turn out to be of relevance for the lattice case.

$\epsilon_{0}^{+}$ and $\epsilon_{0}^{-}$ can be seen to result from
coupling the $S=1/2$ spin either parallel, or antiparallel, with a fully
polarized $T=|\sum_{j}\vt_{j}|=N/2$ set of $\vt$--spins.

All the other $N-1$ states in the $\eta^{z}=(N-1)/2$ subspace belong to the
only other Young tableau which can be constructed with one of the
$\vt$--spins down:
\\
\setlength{\unitlength}{1cm}
\begin{equation}
\begin{picture}(5,3.3)(0,-1)
\thicklines
\put(1.0,0.0){\line(1,0){4.0}}
\put(1.0,1.0){\line(1,0){4.0}}
\put(1.0,-1.0){\line(1,0){1.0}}
\put(1.0,1.0){\line(0,-1){2.0}}
\put(2.0,1.0){\line(0,-1){2.0}}
\put(4.0,1.0){\line(0,-1){1.0}}
\put(5.0,1.0){\line(0,-1){1.0}}
\put(1.5,0.5){\makebox(0,0)[c]{$\uparrow$}}
\put(2.0,0.5){\makebox(2,0)[c]{...}}
\put(4.5,0.5){\makebox(0,0)[c]{$\uparrow$}}
\put(0.5,0.5){\makebox(0,0)[c]{\boldmath {$\Uparrow$}}}
\put(1.5,-0.5){\makebox(0,0)[c]{$\downarrow$}}
\put(3.0,1.3){\oval(3.7,.3)[t]}
\put(3.0,1.8){\makebox(0,0)[c]{\it N-1}}
\end{picture}
\end{equation}
\\
Antisymmetrizing along the column is effectively the same as putting two
$\vt$--spins in a singlet combination, permitting to line up only the
residual $N-1$ spins. The dimensionality of the representation specified
by the YT above is just $d_{1}=N-1$, accounting for all the remaining
$\eta^{z}=(N-1)/2$ states which are thus found to be degenerate. Here we
first see the enormous simplification brought by the $S_N$--symmetry:
spin--rotational symmetry alone would permit the existence of $N$ different
levels corresponding to $\eta=(N-1)/2$ multiplets; $S_N$--invariance tells
us that there are just two levels, corresponding to the two different
possible shapes of Young tableaux. One of these was inherited from the
$\eta^{z}=(N+1)/2$ subspace and corresponds to $T=N/2$; the other is the one
in (11), with a down--spin in the second row, for which $T=(N-2)/2$. --- For
our
particular problem, which has to do with permutational symmetry acting
within $SU(2)$, all YT have at most two rows \cite{group}.

It is crucial to observe that this scheme holds generally: each time when we
step down $\eta^z$, only one new YT appears, namely the one where one box
is removed from the upper row, and a down--spin box is added to the second row.
This corresponds to a new value of the total $\tau$--spin $T$, which is 1
less than the previous lowest $T$--value. Though $\vte=\sum_{j}\vt_{j}$
is not a conserved quantity for the Kondo--term, definite values of $T$ are
unambiguously identified with definite shapes of Young tableaux, $T=(N-2m)/2$
corresponding to a YT with $N-m$ boxes in the upper row.

For $\eta^{z}=(N-2m+1)/2$, most of of the states can be represented by
turning down a spin in the first row of a YT which we had seen previously:
these are stepped--down versions of $\eta>\eta^{z}$ states, corresponding to
levels which had been identified previously. The only new diagrams are
\\
\setlength{\unitlength}{1cm}
\begin{equation}
\begin{picture}(6,4.3)(0,-1.8)
\thicklines
\put(1.0,0.0){\line(1,0){5.0}}
\put(1.0,-1.0){\line(1,0){3.2}}
\put(1.0,1.0){\line(1,0){5.0}}
\put(1.0,1.0){\line(0,-1){2.0}}
\put(2.0,1.0){\line(0,-1){2.0}}
\put(3.2,0.0){\line(0,-1){1.0}}
\put(4.2,0.0){\line(0,-1){1.0}}
\put(5.0,1.0){\line(0,-1){1.0}}
\put(6.0,1.0){\line(0,-1){1.0}}
\put(1.5,0.5){\makebox(0,0)[c]{$\uparrow$}}
\put(3.5,0.5){\makebox(0,0)[c]{...}}
\put(2.6,-0.5){\makebox(0,0)[c]{...}}
\put(5.5,0.5){\makebox(0,0)[c]{$\uparrow$}}
\put(0.5,0.5){\makebox(0,0)[c]{\boldmath {$\Downarrow$}}}
\put(1.5,-0.5){\makebox(0,0)[c]{$\downarrow$}}
\put(3.7,-0.5){\makebox(0,0)[c]{$\downarrow$}}
\put(3.5,1.3){\oval(4.7,.3)[t]}
\put(3.5,1.8){\makebox(0,0)[c]{\it N-m+1}}
\put(2.6,-1.3){\oval(2.9,.3)[b]}
\put(2.6,-1.8){\makebox(0,0)[c]{\it m-1}}
\end{picture}
\end{equation}
\\
associated with combining $T=(N-2m+2)/2$ antiparallel with $S=1/2$, i.e.,
with $\eta=(N-2m+1)/2$, and the diagram with the new YT
\\
\setlength{\unitlength}{1cm}
\begin{equation}
\begin{picture}(6,4.3)(0,-1.8)
\thicklines
\put(1.0,0.0){\line(1,0){5.0}}
\put(1.0,-1.0){\line(1,0){3.2}}
\put(1.0,1.0){\line(1,0){5.0}}
\put(1.0,1.0){\line(0,-1){2.0}}
\put(2.0,1.0){\line(0,-1){2.0}}
\put(3.2,0.0){\line(0,-1){1.0}}
\put(4.2,0.0){\line(0,-1){1.0}}
\put(5.0,1.0){\line(0,-1){1.0}}
\put(6.0,1.0){\line(0,-1){1.0}}
\put(1.5,0.5){\makebox(0,0)[c]{$\uparrow$}}
\put(3.5,0.5){\makebox(0,0)[c]{...}}
\put(2.6,-0.5){\makebox(0,0)[c]{...}}
\put(5.5,0.5){\makebox(0,0)[c]{$\uparrow$}}
\put(0.5,0.5){\makebox(0,0)[c]{\boldmath {$\Uparrow$}}}
\put(1.5,-0.5){\makebox(0,0)[c]{$\downarrow$}}
\put(3.7,-0.5){\makebox(0,0)[c]{$\downarrow$}}
\put(3.5,1.3){\oval(4.7,.3)[t]}
\put(3.5,1.8){\makebox(0,0)[c]{\it N-m}}
\put(2.6,-1.3){\oval(2.9,.3)[b]}
\put(2.6,-1.8){\makebox(0,0)[c]{\it m}}
\end{picture}
\end{equation}
\\
describing the parallel alignment of $T=(N-2m)/2$ total $\vt$--spin with
the $\vs$--spin. This also has $\eta=(N-2m+1)/2$, but belongs to a new
representation of $S_N$, along with the new value of $T$. The diagram
is uniquely associated with the maximum--$\eta^z$ component of the newly
found multiplets which are therefore all degenerate. The degeneracy of the
$T=(N-2m)/2$, $\eta^{z}=(N-2m+1)/2$ level is
\\
\be
d_{m} = C^{N}_{m} - C^{N}_{m-1} = C^{N+1}_{m}\frac{N-2m+1}{N+1}
\ee
\\
where the $C$s denote binomials.

The energies are simply given by
\\
\be
J\vte\cdot\vs = \frac{J}{2} [ \eta(\eta +1)-\frac{3}{4} -T(T+1) ]
\ee
\\
The degeneracy of each energy level is the product of the corresponding
$d_m$, and the $SO(3)$--associated multiplicity $2\eta +1$.

For $N$ even, the procedure finishes with
\\
\setlength{\unitlength}{1cm}
\begin{equation}
\begin{picture}(5,3.3)(0,-1.8)
\thicklines
\put(1.0,0.0){\line(1,0){4.0}}
\put(1.0,-1.0){\line(1,0){4.0}}
\put(1.0,1.0){\line(1,0){4.0}}
\put(1.0,1.0){\line(0,-1){2.0}}
\put(2.0,1.0){\line(0,-1){2.0}}
\put(4.0,1.0){\line(0,-1){2.0}}
\put(5.0,1.0){\line(0,-1){2.0}}
\put(1.5,0.5){\makebox(0,0)[c]{$\uparrow$}}
\put(2.0,0.5){\makebox(2,0)[c]{...}}
\put(2.0,-0.5){\makebox(2,0)[c]{...}}
\put(4.5,0.5){\makebox(0,0)[c]{$\uparrow$}}
\put(0.5,0.5){\makebox(0,0)[c]{\boldmath {$\Uparrow$}}}
\put(1.5,-0.5){\makebox(0,0)[c]{$\downarrow$}}
\put(4.5,-0.5){\makebox(0,0)[c]{$\downarrow$}}
\put(3.0,1.3){\oval(3.7,.3)[t]}
\put(3.0,1.8){\makebox(0,0)[c]{\it N/2}}
\end{picture}
\end{equation}
\\
and a similar diagram with the $S$--spin {\boldmath {$\Downarrow$}}. These
stand for a $d_{N/2}$--dimensional subspace of $\vt$--singlets, leaving the
$S$--spin free to make a disconnected doublet, with energy
$\epsilon_{N/2}^{\pm}=0$. --- For $N$ odd, the last diagram is
\\
\setlength{\unitlength}{1cm}
\begin{equation}
\begin{picture}(6,3.8)(0,-1.8)
\thicklines
\put(1.0,0.0){\line(1,0){5.0}}
\put(1.0,-1.0){\line(1,0){4.0}}
\put(1.0,1.0){\line(1,0){5.0}}
\put(1.0,1.0){\line(0,-1){2.0}}
\put(2.0,1.0){\line(0,-1){2.0}}
\put(4.0,1.0){\line(0,-1){2.0}}
\put(5.0,1.0){\line(0,-1){2.0}}
\put(6.0,1.0){\line(0,-1){1.0}}
\put(1.5,0.5){\makebox(0,0)[c]{$\uparrow$}}
\put(2.0,0.5){\makebox(2,0)[c]{...}}
\put(2.0,-0.5){\makebox(2,0)[c]{...}}
\put(4.5,0.5){\makebox(0,0)[c]{$\uparrow$}}
\put(5.5,0.5){\makebox(0,0)[c]{$\uparrow$}}
\put(0.5,0.5){\makebox(0,0)[c]{\boldmath {$\Downarrow$}}}
\put(1.5,-0.5){\makebox(0,0)[c]{$\downarrow$}}
\put(4.5,-0.5){\makebox(0,0)[c]{$\downarrow$}}
\put(3.5,1.3){\oval(4.7,.3)[t]}
\put(3.5,1.8){\makebox(0,0)[c]{\it (N+1)/2}}
\put(3.0,-1.3){\oval(3.7,.3)[b]}
\put(3.0,-1.8){\makebox(0,0)[c]{\it (N-1)/2}}
\end{picture}
\end{equation}
\\
signifying  an overall singlet $\eta=0$.

In either case, two energy levels belong to each form of the YT with
$0\le{m}\le[(N+1)/2]-1$, and one level goes with $m=[(N+1)/2]$. The total
number of energy levels is $N+1$.
\\
In the presence of an effective field $\omega\ne 0$, the symmetry is reduced
to $S_N$, and $\eta$ is no longer a good quantum number. However, the full
spectrum can still be determined, because the eigenvalue problem is easily
seen to separate into a number of of two--, and one--dimensional problems,
essentially because the $S=1/2$--spin can be stepped only once. Permutation
symmetry requires that only states with the same shape of the YT mix.
Easiest is the case $m=0$, with a YT consisting of a single row of
$\uparrow$--, and $\downarrow$--spins. In the orthonormal basis

\[
\setlength{\unitlength}{1cm}
\begin{picture}(10,2)(0,0)
\thicklines
\put(1.0,0.0){\line(1,0){8.0}}
\put(1.0,1.0){\line(1,0){8.0}}
\put(1.0,1.0){\line(0,-1){1.0}}
\put(2.0,1.0){\line(0,-1){1.0}}
\put(4.0,1.0){\line(0,-1){1.0}}
\put(5.0,1.0){\line(0,-1){1.0}}
\put(6.0,1.0){\line(0,-1){1.0}}
\put(8.0,1.0){\line(0,-1){1.0}}
\put(9.0,1.0){\line(0,-1){1.0}}
\put(1.5,0.5){\makebox(0,0)[c]{$\uparrow$}}
\put(2.0,0.5){\makebox(2,0)[c]{...}}
\put(6.0,0.5){\makebox(2,0)[c]{...}}
\put(4.5,0.5){\makebox(0,0)[c]{$\uparrow$}}
\put(5.5,0.5){\makebox(0,0)[c]{$\downarrow$}}
\put(8.5,0.5){\makebox(0,0)[c]{$\downarrow$}}
\put(9.5,0.5){\makebox(0,0)[c]{=}}
\put(0.5,0.5){\makebox(0,0)[c]{\boldmath {$\Uparrow$}}}
\put(3.0,1.3){\oval(3.7,.3)[t]}
\put(3.0,1.8){\makebox(0,0)[c]{\it N-r}}
\put(7.0,1.3){\oval(3.7,.3)[t]}
\put(7.0,1.8){\makebox(0,0)[c]{\it r}}
\end{picture}
\]

\be
= \frac{1}{\sqrt{C^{N}_{r}}}\sum_{j_{1}<...<j_{r}}|\Uparrow\rangle|\uparrow...
\downarrow...\downarrow...\uparrow\rangle
\ee
\\
and
\\
\[
\setlength{\unitlength}{1cm}
\begin{picture}(10,2)(0,0)
\thicklines
\put(1.0,0.0){\line(1,0){8.0}}
\put(1.0,1.0){\line(1,0){8.0}}
\put(1.0,1.0){\line(0,-1){1.0}}
\put(2.0,1.0){\line(0,-1){1.0}}
\put(4.0,1.0){\line(0,-1){1.0}}
\put(5.0,1.0){\line(0,-1){1.0}}
\put(6.0,1.0){\line(0,-1){1.0}}
\put(8.0,1.0){\line(0,-1){1.0}}
\put(9.0,1.0){\line(0,-1){1.0}}
\put(1.5,0.5){\makebox(0,0)[c]{$\uparrow$}}
\put(2.0,0.5){\makebox(2,0)[c]{...}}
\put(6.0,0.5){\makebox(2,0)[c]{...}}
\put(4.5,0.5){\makebox(0,0)[c]{$\uparrow$}}
\put(5.5,0.5){\makebox(0,0)[c]{$\downarrow$}}
\put(8.5,0.5){\makebox(0,0)[c]{$\downarrow$}}
\put(9.5,0.5){\makebox(0,0)[c]{=}}
\put(0.5,0.5){\makebox(0,0)[c]{\boldmath {$\Downarrow$}}}
\put(3.0,1.3){\oval(3.7,.3)[t]}
\put(3.0,1.8){\makebox(0,0)[c]{\it N-r+1}}
\put(7.0,1.3){\oval(3.7,.3)[t]}
\put(7.0,1.8){\makebox(0,0)[c]{\it r-1}}
\end{picture}
\]

\be
= \frac{1}{\sqrt{C^{N}_{r-1}}}\sum_{j_{1}<...<j_{r-1}}|\Downarrow\rangle
|\uparrow...\downarrow...\downarrow...\uparrow\rangle
\ee
\\
$H_{MF}$ is represented by the matrix
\bea
\left( \begin{array}{cc}
(1/4N)(J-2\omega)(N-2r) & (J/2N)\sqrt{r(N-r+1)} \\\\
(J/2N)\sqrt{r(N-r+1)} & -(1/4N)(J+2\omega)[N-2(r-1)]
\end{array} \right)
\eea
\\
The ground state is now non--degenerate, corresponding to $r=1$, and a
maximum polarization along the effective field. Note that while, in the
absence of an external field term, the pseudospin model gives only a very
poor imititation of the highly non--trivial quantum--mechanical ground state
of the electronic overscreened model, we have much less reason to doubt the
validity of the picture obtained from the necklace model for moderately
strong external fields. Since in a lattice, intersite interactions amount to
a field acting at any particular Kondo site, herein lies our hope that
the multichannel necklace model can give some guidance as to the behaviour
of the multichannel Kondo lattice. We have already quoted arguments
\cite{{ATs},{IJW}} showing that intersite interactions can make the system
turn towards a strong--coupling fixed point: this kind of behaviour can
be imitated by a necklace--type model.

Similarly, for states with $m$ $\downarrow$--spins in the second row of
the YT, the two--dimensional subspace is spanned by
\\
\setlength{\unitlength}{1cm}
\begin{equation}
\begin{picture}(9.25,4.3)(0,-1.8)
\thicklines
\put(1.0,0.0){\line(1,0){8.25}}
\put(1.0,-1.0){\line(1,0){3.75}}
\put(1.0,1.0){\line(1,0){8.25}}
\put(1.0,1.0){\line(0,-1){2.0}}
\put(2.0,1.0){\line(0,-1){2.0}}
\put(4.25,1.0){\line(0,-1){1.0}}
\put(5.25,1.0){\line(0,-1){1.0}}
\put(6.25,1.0){\line(0,-1){1.0}}
\put(8.25,1.0){\line(0,-1){1.0}}
\put(9.25,1.0){\line(0,-1){1.0}}
\put(3.75,0.0){\line(0,-1){1.0}}
\put(4.75,0.0){\line(0,-1){1.0}}
\put(1.5,0.5){\makebox(0,0)[c]{$\uparrow$}}
\put(2.125,0.5){\makebox(2,0)[c]{...}}
\put(1.875,-0.5){\makebox(2,0)[c]{...}}
\put(6.25,0.5){\makebox(2,0)[c]{...}}
\put(4.75,0.5){\makebox(0,0)[c]{$\uparrow$}}
\put(5.75,0.5){\makebox(0,0)[c]{$\downarrow$}}
\put(8.75,0.5){\makebox(0,0)[c]{$\downarrow$}}
\put(0.5,0.5){\makebox(0,0)[c]{\boldmath {$\Uparrow$}}}
\put(1.5,-0.5){\makebox(0,0)[c]{$\downarrow$}}
\put(4.25,-0.5){\makebox(0,0)[c]{$\downarrow$}}
\put(3.125,1.3){\oval(3.8,.3)[t]}
\put(3.125,1.8){\makebox(0,0)[c]{\it N-m-r}}
\put(2.875,-1.3){\oval(3.3,.3)[b]}
\put(2.875,-1.8){\makebox(0,0)[c]{\it m}}
\put(7.25,1.3){\oval(3.7,.3)[t]}
\put(7.25,1.8){\makebox(0,0)[c]{\it r}}
\end{picture}
\end{equation}
\\
and
\\
\begin{equation}
\begin{picture}(9.25,4.3)(0,-1.8)
\thicklines
\put(1.0,0.0){\line(1,0){8.25}}
\put(1.0,-1.0){\line(1,0){3.75}}
\put(1.0,1.0){\line(1,0){8.25}}
\put(1.0,1.0){\line(0,-1){2.0}}
\put(2.0,1.0){\line(0,-1){2.0}}
\put(4.25,1.0){\line(0,-1){1.0}}
\put(5.25,1.0){\line(0,-1){1.0}}
\put(6.25,1.0){\line(0,-1){1.0}}
\put(8.25,1.0){\line(0,-1){1.0}}
\put(9.25,1.0){\line(0,-1){1.0}}
\put(3.75,0.0){\line(0,-1){1.0}}
\put(4.75,0.0){\line(0,-1){1.0}}
\put(1.5,0.5){\makebox(0,0)[c]{$\uparrow$}}
\put(2.125,0.5){\makebox(2,0)[c]{...}}
\put(1.875,-0.5){\makebox(2,0)[c]{...}}
\put(6.25,0.5){\makebox(2,0)[c]{...}}
\put(4.75,0.5){\makebox(0,0)[c]{$\uparrow$}}
\put(5.75,0.5){\makebox(0,0)[c]{$\downarrow$}}
\put(8.75,0.5){\makebox(0,0)[c]{$\downarrow$}}
\put(0.5,0.5){\makebox(0,0)[c]{\boldmath {$\Downarrow$}}}
\put(1.5,-0.5){\makebox(0,0)[c]{$\downarrow$}}
\put(4.25,-0.5){\makebox(0,0)[c]{$\downarrow$}}
\put(3.125,1.3){\oval(3.8,.3)[t]}
\put(3.125,1.8){\makebox(0,0)[c]{\it N-m-r+1}}
\put(2.875,-1.3){\oval(3.3,.3)[b]}
\put(2.875,-1.8){\makebox(0,0)[c]{\it m}}
\put(7.25,1.3){\oval(3.7,.3)[t]}
\put(7.25,1.8){\makebox(0,0)[c]{\it r-1}}
\end{picture}
\end{equation}
\\
The eigenvalue problem is similar to that of $m=0$, only we have to
remember that in a YT, the states are antisymmetrized along the columns, so
the symmetrization of $\uparrow$-- and $\downarrow$--spins along the first
row is not allowed to bring $\downarrow$--spins into the first $m$ boxes.
Effectively, in (20), in the off--diagonal elements $N$ must be replaced by
$N-2m$, giving the matrix
\\
\bea
\left( \begin{array}{cc}
(1/4N)(J-2\omega)[N-2(m+r)] & (J/2N)\sqrt{r(N-2m-r+1)} \\\\
(J/2N)\sqrt{r(N-2m-r+1)} & -(1/4N)(J+2\omega)[N-2(m+r-1)]
\end{array} \right)
\eea
\\
The solution of the eigenvalue problem would be trivial to write down, but
we do not need the lengthy expressions. First of all, we are interested in
the mean field result for the magnetic transition temperature $T_N$, for
which linearized eigenvalues
\\
\be
\lambda^{+}_{mr} = \frac{J}{4N}(N-2m) -
\frac{(N-2m)(N+1-2m-2r)}{2N(N+1-2m)}\omega
\ee
\\
and
\\
\be
\lambda^{-}_{mr} = -\frac{J}{4N}(N+2-2m) -
\frac{(N+2-2m)(N+1-2m-2r)}{2N(N+1-2m)}\omega
\ee
\\
are sufficient. We also need $\lr{\tau}$ for the corresponding eigenstates
\\
\be
\tau^{+}_{mr} = \frac{(N-2m)(N+1-2m-2r)}{2N(N+1-2m)} -
\frac{4r\omega}{JN} \frac{N+1-2m-r}{(N+1-2m)^{3}}
\ee
\\
and
\\
\be
\tau^{-}_{mr} = \frac{(N+2-2m)(N+1-2m-2r)}{2N(N+1-2m)} +
\frac{4r\omega}{JN} \frac{N+1-2m-r}{(N+1-2m)^{3}}
\ee
\\
In the linearized self--consistency equation for $\lr{\tau}$, the denominator
is just the $\omega=0$ value of the single--site partition function
\\
\be
Z_{0} = \sum_{m=0}^{[\frac{N+1}{2}]-1} d_{m}
\left[ (N-2m+2)e^{-J(N-2m)/JNT_{N}} +
(N-2m)e^{J(N-2m+2)/4NT_{N}} \right]
\ee
\\
while the numerator is the suitably weighted sum of (26) and (27)
\\
\be
\lr{\tau} = \frac{1}{Z_{0}} \sum_{m=0}^{[\frac{N+1}{2}]-1} d_{m}
\left[ \sum_{r=1}^{N-2m} e^{-\lambda^{-}_{mr}/T_{N}} \tau^{-}_{mr} +
\sum_{r=0}^{N-2m+1} e^{-\lambda^{+}_{mr}/T_{N}} \tau^{+}_{mr} \right]
\ee
\\
Actually, a slight complication has to be dealt with before arriving at (29).
In the second $r$--sum, the terms $1\le r\le N-2m$ arise from the
two--dimensional eigenvalue problem (23). However, the state with $r=0$,
and $\vs$--spin $\Uparrow$, is not coupled to any other state, it presents
a one--dimensional eigenvalue problem. Similarly for $r=N-2m+1$, and the
$\vs$--spin $\Downarrow$. These are actually the $\eta^{z}={\pm}(N-2m+1)/2$
components of the corresponding $\eta=(N-2m+1)/2$ multiplets, and are thus
eigenstates by construction. It is, however, easy to check
that these cases are accounted for by extending the second sum in (29) to
include $r=0$, and $r=N-2m+1$.

Though the $r$--sums in (29) are easily done (with the result that $\lr{\tau}$
disappears from the equation), the $m$--summation is not, and we
did not find any transparent from into which (29) could be cast. Essential
simplification can be achieved only for the limiting case $J/W\to\infty$
when it is sufficient to consider the action of $H_{MF}$ within the subspace
of the $\eta=(N-1)/2$ Kondo ground states associated with the diagrams shown
in (8), and (9). From (29), we find
\\
\be
\lim_{J/W\to\infty} T_{N} = \frac{W}{6} \frac{(N-1)(N+2)^{2}}{N^{2}(N+1)}
\ee
\\

The above result can be easily interpreted by noticing that within the
Kondo ground state set of states $\vt_j$ acts like
\\
\be
\vt_{j} \Rightarrow \frac{N+2}{N(N+1)}\veta
\ee
\\
independently of $j$. Thus, in the restricted Hilbert space formed by taking
the direct product of the Kondo ground state sets for each site, the
intersite coupling term in (3) can be exactly represented by the effective
hamiltonian
\\
\be
h_{eff} = W\frac{(N+2)^{2}}{N(N+1)^{2}} \sum_{\lr{{\bf i},{\bf j}}}
[\eta_{\bf i}^{x}\eta_{\bf j}^{x}+\eta_{\bf i}^{y}\eta_{\bf j}^{y}]
\ee
\\
(30) is just the mean--field solution for (32). Note that it gives 0 for
$N=1$; this is in accordance with Doniach's finding \cite{Do77} of a
transition to the fully Kondo--compensated state at $J/W=1$ for the original
necklace model.

For general values of $J/W$, we solved (29) numerically. Fig. 1 shows the
coupling dependence of $T_{N}$, for several values of $N$. On the horizontal
axis, we have chosen the variable $J/(J+W)$, so as to be able to include the
asymptotic regime where (30) becomes valid. (Since in the MFA, the
coordination number $z$ enters only through $zW$, we have simply taken
$z=2$).

At $J=0$, the system separates
into $N$ independent channels, disconnected from the $\vs$--spins. Each set
of $\vt$--spins orders at the same mean--field temperature $W/2N$. This point
is clearly pathological; our interest lies in the behaviour at intermediate
(and in any case, non--zero) values of $J/W$.

First of all, one should note that, with the exception of $N=2$, all curves
show an overall rising tendency with increasing $J/W$: the asymptotic value
(30) is lying significantly higher than the low-$J$ value of $T_N$.
Remarkably, the Kondo effect does not seem to compete with pseudopsin
ordering but rather to assist
it ! We can understand this tendency by remembering that the ordering is
that of composite spins [most clearly seen in the large--$J$ expression (32)]
which arise from glueing the $\vt$--spins together. In the starting
hamiltonian (3), there is no direct interchannel coupling: the effective
ferromagnetic coupling between the channels is done by the Kondo coupling.
So, in the multi--channel case, the Kondo coupling is effectively creating
the objects which  subsequently order.

A subtler feature of our $T_N$ curves is a maximum
at some $J/W$ value which appears to increase with $N$. For a better view, we
give a blow--up of the relevant region of the phase diagram in Fig. 2. The
broad, relatively
low maximum is most discernable for $N=2$, but it apparently exists (though
quickly becoming quite inconspicuous) for any finite $N$.

We have also done the numerical solution of the self--consistency equation
at arbitary temperatures, i.e., at finite effective field strengths. We
found that the temperature dependence of the order parameter is of the
form usually found in similar mean--field solutions, so we renounce
including them here.

In our pseudospin model, the ordering which sets in at $T_N$, is (depending
on the sign of $W$), either a ferromagnetic, or a two--sublattice
antiferromagnetic, ordering of the $\vt$--spins. Via the Kondo--coupling,
this induces a similar, but oppositely polarized, ordering of the
$\vs$--spins. It depends on the nature of the underlying overscreened Kondo
model, what the physical meaning of this order is. In Cox's \cite{Cox87}
model of U--compounds, it is primarily the ordering of electric quadrupole
moments. The possibility that actual spin magnetization is a secondary order
parameter arising from the mixing--in of higher--lying ionic levels, is an
attractive possibility to account for the smallness of the ordered magnetic
moment of some heavy fermion magnets \cite{LiCo}. It remains an open question
whether quadrupolar Kondo effect can assist superconductivity \cite{Cox1}.

\section{\normalsize DISCUSSION AND SUMMARY}

The crucial question of the theory of Kondo lattices is to what extent can
the results for a single Kondo impurity be taken as a guide for the
behaviour of periodic systems ? The answer must certainly depend on the size
of the localized spin $S$, and the number of conduction electron channels
$N$. We are uncertain about the answer even in the relatively simple cases
of the underscreened ($S>N/2$), and exactly screened ($S=N/2$), Kondo
lattices. The situation is quite obscure for the overscreened ($S<N/2$)
Kondo lattice since it is difficult to infer how the subtle, and puzzling,
features revealed by the solution of the single--ion multi--channel Kondo
problem should manifest themselves in the lattice case. The available
arguments \cite{{ATs},{IJW}} suggest that switching on intersite interactions
must have a drastic effect.

Inspired by the fact that valuable insight into the behaviour of the
$S=1/2$, $N=1$ Kondo lattice has been gained from studying the Kondo
necklace model introduced by Doniach \cite{Do77}, we set out to define and
investigate similar pseudospin models corresponding to a variety of Kondo
lattices. Following our earlier study of underscreened Kondo necklaces
\cite{FK}, here we introduced the $N$--channel, $S=1/2$, overscreened
necklace model (3). In the
spirit of necklace models, the $N$ screening channels of the underlying
Kondo lattice problem are represented by $N$ sets of pseudospins $\vt_j$.
The low--lying electron--hole excitations of the Fermi sea, and hence the
possibility of a ``true'' Kondo effect, are thereby lost. Of the conduction
electrons, only their spin degrees of freedom are retained: the pseudospin
model still incorporates the competing tendencies of a Kondo--like spin
compensation, and magnetic ordering. Intersite coupling is mediated by the
propagating character of the pseudospins, described by an $x-y$ coupling
term in (3).

We studied the phase diagram of the model in the same kind of mean--field
approximation (MFA) as that used by Doniach \cite{Do77}. Setting up the
self--consistency equation requires the diagonalization of the single--site
effective hamiltonian (4) in a 2$^{N+1}$--dimensional Hilbert space: this
is made possible by exploiting the invariance under permutations of the
channel indices. We found that the low--temperature phase is always ordered;
for $W>0$, and bipartite lattices, this is just the N\'eel order of the
$\vt$--spins, accompanied by the oppositely polarized ordering pattern of
the $\vs$--spins. All the channels are equally polarized (5), corresponding
to the well--known tendency that the Kondo--term mediates an effective
ferromagnetic coupling between the different channels.

The nearly parallel alignment of the pseudospins results in a remarkable
effect: for $N\ge 3$, the ordering tendency is actually enhanced with
increasing Kondo coupling $J$ ! This can be understood as arising from the
fact that overscreening builds up composite pseudospin objects which are
larger than the ``naked'' original spin $S$. Their coupling, mediated by
the pseudospin $x-y$ term, turns out to be sufficiently effective to increase
the N\'eel temperature $T_N$.

In the terms of the model (3), the ordering is of the easy--plane type.
This is most clearly born out by the exact form (32) of the effective
hamiltonian which becomes valid in the limit of infinite Kondo coupling.
Introducing the $x-y$ form of the intersite $\vt$--coupling has
destroyed the spin--rotational invariance of the underlying electronic
hamiltonian, which is still shown by the isotropic Kondo--coupling. In
this sense, the necklace models are in a universality class different from
that of the Kondo lattices \cite{SaSo}, and this has to be kept in mind
when trying to transfer results from one class of models to the other.

The low--temperature state of the model can be visualized as the ordering of
composite $|\veta|=(N-1)/2$ spins. This apparently corresponds to the
strong--coupling behaviour of the overcompensated Kondo--site, and not to
its intermediate--coupling fixed point with the strange, quantum--mechanical
non--integer zero point degeneracy \cite{{WTs},{AL91}}.  The conjecture
\cite{GAC93} that the overscreened lattice might show a corresponding
exotic phase is certainly exciting but this is not born out by our study
of a drastically simplified model. On the other hand, we can relate our
findings to independent arguments \cite{{ATs},{IJW}} suggesting that
intersite interactions are likely to drive the system towards the
strong--coupling regime.

Noting that the multi--channel Kondo problem can be used to model a
variety of systems \cite{{Z},{Cox87},{Sea},{KMF}}, the physical nature
of the predicted order can also be different from case to case. Thinking of
potential applications we have, however, primarily the suggested quadrupolar
ordering of U--based systems in mind \cite{{Cox87},{LiCo}}.

To summarize, we introduced the $N$--channel Kondo necklace model for a
preliminary study of the nature of the collective spin state in
overscreened Kondo lattices. Our findings indicate that intersite
interactions drastically change the state of the Kondo centres, lifting
the ground state degeneracy, and inducing (pseudo)spin ordering at low
temperatures.

\vspace{.8cm}

\centerline{ACKNOWLEDGEMENTS}
\vspace{.4cm}
The authors wish to express their gratitude
to the International Centre for Theoretical Physics for financial support,
hospitality, and an encouraging scientific atmosphere. P.F. is indebted also
to SISSA (Trieste) for the hospitality extended to him. Useful discussions
with, and helpful advice from, A. Zawadowski and D.L. Cox are gratefully
acknowledged.

\newpage
\parindent 0pt
{\bf Figure Captions}
\\\\\
{\em Fig. 1} The dependence of the N\'eel temperature $T_N$ on the
dimensionless
Kondo coupling $J/W$, for $N$=2,3,4,5, and 10, and $z=2$. Choosing the
variable $J/(J+W)$ on the horizontal
axis compresses the half--axis $0\le J/W<\infty$ into a finite interval.
\\\
{\em Fig. 2} Enlargement of part of Fig. 1 to show more clearly the maxima
on the $T_N$ {\em vs} $J/(J+W)$ curves. The positions of the maxima are
indicated by diamonds. The dotted line is a guide to the eye; it indicates
that the maximum shifts to increasingly large values of $J/W$ with
increasing $N$.

\end{document}